# Does Regression Approximate the Influence of the Covariates or Just Measurement Errors? A Model Validity Test.

Alexander Kukush[a] and Igor Mandel[b]

[a]Faculty of Mechanics and Mathematics, Taras Shevchenko National University of Kyiv, Kyiv, Ukraine; [b]Redviser Inc., New York, NY, USA

**Abstract**

A criterion is proposed for testing hypothesis about the nature of the error variance in the dependent variable in linear model, which separates correctly and incorrectly specified models. In the former only measurement errors determine the variance (i.e., dependent variable is correctly explained by independent ones, up to measurement errors), while in the latter the model lacks some independent covariates (or has nonlinear structure). The proposed MEM-V (Measurement Error Model Validity) test checks the validity of the model when both dependent and independent covariates are measured with errors. The criterion has asymptotic character, but numerical simulations outlined approximate boundaries where estimates make sense. A practical example of the test's implementation is discussed in detail; it shows test's ability to detect wrong specification even in seemingly perfect models. This type of relation between measurement errors and model's specification has not been studied before, and the proposed criterion may stimulate future research in this important area.

## 1. INTRODUCTION

As well-known, the origination of the least squares (LS) method of parameter estimation by A. Legendre, F. Gauss and others was inspired by the problem of measurement errors, mainly in astronomy, and was intended exclusively to address that issue. But further development of statistics left those ideas far behind, and regression technique based on LS (and later on other principles, like maximal likelihood) became a staple approach for any analysis, where certain relations between variables are assumed and the purpose was to estimate the values of the "real parameters" governing those relations. After many decades of development and countless number of studies, some questions related to this approach remain debatable, even in the simplest linear case. The uncertainty and misunderstanding are sneaking between several interrelated issues, such as the nature of variables involved and the actual processes behind the scene; type of measurement errors for each covariates and their interplay; causal or just "statistically significant" meaning of the parameters and others.

Consider a typical model, so often to be found in different versions in statistical literature:

$$y = c^T x + e, \qquad (1)$$

where $x$ is a column vector of certain covariates (explanatory or independent variables); $y$ is the response (dependent) variable; $c$ is a column vector of the "actual parameters" to be estimated; $e$ is the error term. There are many interpretations of those linear models and, respectively, ways to make estimates. Possible, but not all combinations of different assumptions about four elements of the models, $y, x, c, e$, each of which is needed for correct model estimation, could be summarized in table 1. It does not account for conditions like "endogenous" and "exogenous"; "latent and observable"; normal errors distributions or not; homoskedasticity and some others. Even without those the number of assumptions to be made is too high.

Just one glance on the table shows how much idealization the current statistics contain. One cannot realistically assume that anything could be measured without the errors – and yet this assumption prevails in all statistical literature, it goes usually without saying in standard statistical textbooks or even in specialized volumes for linear models like [1].

Similarly, the distinction between "causal and non-causal" coefficients is seemingly non-existent in literature: statistical books and articles usually do not use causality terminology, while very developed causality modeling works ([2,3] and others) tend to consider the coefficients $c$ obtained by usual regression as causal only because those regressions are imbedded into special framework (direct acyclic graph, DAG), while in essence this inclusion does not change the nature of the coefficients by itself, what was shown in detail in [4,5]. Just recently some proposals about distinction of the causal and non-causal variables have appeared [5-7] – and yet, they do not work in combination with measurement error theory (to leave aside that they are far from perfect).

Table 1. Different assumptions about elements of the linear model (1),
needed for correct estimation of the parameters

|  |  |  |  | Independent covariates x | | | |
|---|---|---|---|---|---|---|---|
|  |  |  |  | Measured without errors | | Measured with errors | |
|  |  |  |  | **Coefficients c** | | **Coefficients c** | |
|  |  |  |  | Non-causal | Causal | Non-causal | Causal |
|  |  |  |  | a | b | c | d |
| **Dependent variable y** | Measured without errors | Simple e | 1 | 1a | 1b | 1c | 1d |
|  |  | Composite e | 2 | 2a | 2b | 2c | 2d |
|  | Measured with errors | Simple e | 3 | 3a | 3b | 3c | 3d |
|  |  | Composite e | 4 | 4a | 4b | 4c | 4d |

A comment is needed here though. The very term "causal" as used in statistical modeling could be very controversial (see many definitions and references in [2,4,5]), but on intuitive level the "causes" could be divided in two types: those which generate outcome directly, by the "law of nature", like force generating the spring extension in Hooke's law, and those which generate outcome probabilistically, via different intermediators, like tax reduction stimulating the growth of business activity months later after thousands of channels will be found to use newly opened benefits. Needless to say, the causes of the second type dominate the world and are the main concern of the theoretical literature. However, the distinction between the two is not easy to grasp with solely statistical tools at hands, and the measurement errors play a crucial role here. Imagine, that applied force *F* in Hooke's law *F=kx* (hanging weights, originally) was measured very badly – with only large warehouse scale available instead of the appropriate small one. In that case the independent covariate has high measurement error which results in the so-called *attenuation effect* (when ordinary LS estimate $\hat{c}$ is closer to zero than the true vector *c* [8]), which means that no effect, causal or not, is to be revealed to begin with. The coefficient *k* will never be correctly estimated, and the very law will never be discovered -neither by R. Hooke nor by anyone else. In that sense the measurement problem lies at the heart of physics and other nature sciences; relation to measurement separates them from other forms of culture [21].

The high measurement error will be seen numerically as a low level of approximation of the extension *x* by force *F*. But this is exactly what is observed in thousands of applications, when causes of the second type, not those natural ones, presumably play role. How to distinguish the deviation of the response *y* from its prediction by *x* due to just measurement errors vs. the deviation due to "legitimate" reasons? How to separate confidently two statements: "*economic growth is determined not only by tax reduction, but by many other factors as well*" and "*economic growth is a direct function of the tax reduction only, but both are measured badly, and for that reason we do not see the perfect approximation*"? This is a topic of the paper, and in this form, to the best of our knowledge, it was not discussed in the literature yet, despite of its obvious importance.

In the light of those considerations, let stop by on the interpretation of the so-called random term *e* in (1) and in table 1. Technically, it looks like a value needed to be added to the deterministic term of the equation $c^T x$ in order to be equal the *y* value precisely (or, in backward interpretation, as "residual" when Y is already estimated). Neither the nature of that addition nor the reason of its appearance is usually considered. It is just assumed, that some difference always exists and one needs to minimize it (the whole idea of LS and all other methods is exactly that). Typically, it is considered as some kind of random variables generated by each observation with convenient features, like this:$e_i$ "*…are independent and identically distributed random variables with zero mean and constant variance*" [9, p.3].

This type of treatment of those "residuals" holds a key question about their nature unanswered. Say, if two covariates should determine *y* (in the sense of strict causes like in Hooke's law), but we have in a model just one of them, then the interpretation and assumption about i.i.d. and, perhaps, equal variances (homoskedasticity) will not change. Equally, if suddenly the third covariate, collinear to one of those, included in a model (with spoiling effect) – the interpretation, again, will be as it is. But the crucial difference is that only when two causal variables are in a model, *e* becomes a purely error of measurement, while in any misspecification it is not true. We, respectively, call errors terms in table 1*simple* when they are assumed as reflecting measurement errors only, and *composite*, when they are assumed to represent also model misspecification. A typical confusion is that errors are considered as i.i.d. random variables regardless on the fact they are simple or composite, while in the latter scenario the randomness should be violated and/or have another explanation (even in the case of symmetric or normal distribution of the known measurable errors).

Now, we can have a look at the table 1 again. Assumptions described in columns a, b and rows 1,2 may be valid only in imagination of mathematicians. Non-causal models (columns a and c), in theory, should, in theory, interest no one: in the better world researches should always assume that they are busy with the real causes, not with fictitious entities, and use respective technical tools, whatever good or bad. But in practice, only those unrealistic conditions are investigated with the whole strength possible. Perhaps, 90% of all statistical literature considers cases 1a and 3a (much rarer), usually not naming them as such. Vast literature about measurement errors (see recent review [8], thesaurus article [10], monograph [11], etc.) in fact does not distinguish between causal and not-causal covariates and between model misspecification and pure random errors, i.e., it is concentrated in cells 3c, 3d, 4c, 4d, but lumping them into one block. Ideally, the only worthwhile models should belong to the class 4d, i.e., they should distinguish causal and non-casual covariates in situations with measurement errors and misspecification. But in fact, they represent a small fraction of all works. This bias strongly contributes into the roots of the actively discussed now "reproducibility crisis" [12] and "statistical significance crisis" [13, 14]. A lot of work for investigating different conditions in a sense of the table 1 lies ahead, in our opinion.

In this paper, we try to deal with just one question in a realm of the ones considered above: in what case one may say with confidence, that the built model does not reflect just measurement errors, but measure something more substantial. Or, in other words, having any idea about possible level of the measurement errors in the data – can one say, that the model captures strong (not necessarily causal) relations between variables, which are not hidden by measurement errors? So, the paper considers situations in cells 2c, 2d, 4c, 4d, i.e., belongs to a rare type of studies separating composite error terms from just random, without, however, special stress for causality. Such situations are technically related to the so-called *equation error model* (see Section 1.5 in [11]).

The proposed criterion we named MEM-V (Measurement Error Model Validity) test, for it distinguishes valid (i.e. complete) models where only inevitable measurement errors play role, and invalid (incomplete) models, where hidden covariates (not included in a model) affect the variance of the independent covariate. Moreover, even in complete model the test rejects the validity hypothesis when assumed errors in covariates are too high. The test allows, more specifically, answering the following question: is the quality of the model (say, coefficient of determination or another indicator) higher than the level which could be obtained just by approximating the measurement errors? If yes, then the model makes sense; if no - it does not, regardless of its goodness-of-fit level. This type of criterion could be very useful in situations, where a researcher has a plausible idea about the level of measurement errors either for *y* or for *x,* or just knows for sure about it (when, for example, the precision of the measurement devices is accessible). In certain sense it inverses the usual formulation of the errors in variables problem: how to adjust the coefficients of regression in order to account for measurement errors of different types [8, 10]. In this latter scenario one gets the estimates with wider confidence intervals, but still believing that she found "something". In the former one knows that nothing good could be expected if the model covers just errors and (erroneously) the hidden factors and nothing more.

The rest of the article is organized as follows. Section 2 describes a specific case of model (1), where the covariates *x* are observed with additive measurement error; in our approach, the variance $\sigma^2 = \boldsymbol{Var}(e)$ is a parameter of interest and *c* plays a role of nuisance parameter (unlike under usual treatment); based on the assumption of the independent observations, we constructed a consistent and asymptotically normal estimator of $\sigma^2$, that yields a one-sided Wald test which serves as a desired criterion for separation of the correctly and incorrectly specified models. Section 3 contains results of the simulation experiment. Section 4 demonstrates how to implement the test on the real data and presents some non-trivial results; Conclusion summarizes the main results and outlines the area of the future research. Appendix 1 contains proofs of the asymptotic properties of the estimators and Appendix 2 studies the behavior of the test for large sample size and growing level of the presumed measurement errors.

All the vectors in the paper are column ones. Abbreviation r.v. means random variable; $\boldsymbol{Var}$ stands for the variance of a r.v. and $\boldsymbol{Cov}$ for the variance-covariance matrix of a centered random vector.

## 2. HYROTHESIS TESTING CONCERNINGTHE ERROR VARIANCE

### 2.1. Basic Regression Model

We deal with a linear functional errors-in variables model

$$y = c^T x + e, \qquad (2)$$

$$w = x + \delta. \qquad (3)$$

Here *x* is a nonrandom vector of covariates valued in $\boldsymbol{R}^m$; *y* is an observable response variable; $\delta$ is a vector of measurement errors and *e* is another error term that may appear not only as a result of measurement, but as an effect of misspecification (e.g., lack or of covariates actually affecting *y*); instead of true covariates *x*, we observe a surrogate data *w*; $c \in \boldsymbol{R}^m$ is an unknown regression parameter.

Concerning error terms, we assume the following.

(i) Error *e* in response is a centered r.v. and measurement error $\delta$ in covariatesis a centered random vector, moreover *e* and $\delta$ are independent with finite 2nd moments.

(ii) The error variance $\sigma^2 := \boldsymbol{Var}(e)$ is not vanishing and unknown, while the measurement error covariance matrix $S := \boldsymbol{Cov}(\delta)$ is known, where *S* is a positive semidefinite matrix (see Remarks below for more detail).

**Remark 1.** It is allowed that some components $\delta^{(j)}$ of $\delta$ have zero variance. Thus, a subgroup of covariates can be observed without measurement error.

**Remark 2.** In practice, sometimes *S* can be reliably estimated by repeated measurements of covariates. Another possible case is as follows: components of $\delta$ are independent and $S = diag(\tau_1^2, \dots, \tau_m^2)$ with $\tau_j^2 = \boldsymbol{Var}(\delta^{(j)})$ known from the passport information of measurement devices. In case $S = 0$, we have ordinary linear regression (2) with both observable *y* and *x*.

**Remark 3.** Other possible cases of available information concerning $\sigma^2$ and *S* will be indicated in Conclusion. Note that some information about the error covariance structure should be available, otherwise the model (2) - (3) is not identifiable, and in that case even for Gaussian errors it is impossible to estimate all the model parameters consistently as the sample size tends to infinity [11].

Individual observations in a data set are modeled as independent copies of the basic model

$$y_i = c^T x_i + e_i, \tag{4}$$

$$w_i = x_i + \delta_i, \quad i = 1, \ldots, n. \tag{5}$$

Here $x_1, \ldots, x_n$ are nonrandom and unknown vectors; $\{e_i, i \geq 1\}$ are independent i.i.d. random variables distributed as $e$ in the model (2-3), $\{\delta_i, i \geq 1\}$ are i.i.d. random vectors distributed as $\delta$, and the two sequences are assumed mutually independent. Based on observations

$$(y_1, w_1), \ldots, (y_n, w_n), \tag{6}$$

we want to estimate $\sigma^2 = \mathbf{Var}(e)$ and test a hypothesis that its value does not exceed a fixed level.

## 2.2. Consistent Estimators of Model Parameters

As a consistent estimator of $c$, we use the so-called *adjusted least squares* (ALS) estimator [11]:

$$\hat{c} = (\overline{ww^T} - S)^+ \overline{wy}. \tag{7}$$

Hereafter bar stands for averaging over the whole sample, in particular $\overline{ww^T} := \frac{1}{n}\sum_{i=1}^{n} w_i w_i^T, \overline{wy} = \frac{1}{n}\sum_{i=1}^{n} w_i y_i$; $D^+$ denotes the pseudo-inverse matrix of a matrix $D$. If the matrix $\overline{ww^T} - S$ is nonsingular, then (7) takes a form

$$\hat{c} = (\overline{ww^T} - S)^{-1} \overline{wy}. \tag{8}$$

Under mild assumptions this happens with probability tending to 1 as $n \to \infty$ (see statement 1 of Theorem in Appendix 1).

A consistent estimator $\tilde{\sigma}^2$ of $\sigma^2$ is based on the following computation:

$$E(y - c^T w)^2 = E(e - c^T \delta)^2 = \sigma^2 + c^T S c. \tag{9}$$

Thus, $\tilde{\sigma}^2$ is expressed through the residual sum of squares $RSS := \sum_{i=1}^{n}(y_i - \hat{c}^T w_i)^2$,

$$\tilde{\sigma}^2 := n^{-1} RSS - \hat{c}^T S \hat{c}. \tag{10}$$

Introduce estimating functions

$$s_c = s_c(y, w; c) = wy - (ww^T - S)c, \tag{11}$$

$$s_{\sigma^2} = s_{\sigma^2}(y, w; c, \sigma^2) = (y - c^T w)^2 - c^T S c - \sigma^2. \tag{12}$$

With probability tending to 1 as $n \to \infty$, the pair $(\hat{c}; \tilde{\sigma}^2)$ is a solution to estimating equations

$$\sum_{i=1}^{n} s_c(y_i, w_i; c) = 0, \quad \sum_{i=1}^{n} s_{\sigma^2}(y_i, w_i; c, \sigma^2) = 0, \quad c \in \mathbf{R}^m, \quad \sigma^2 \geq 0. \tag{13}$$

The estimator (10) is consistent, i.e. $\tilde{\sigma}^2$ converges in probability to $\sigma^2 > 0$ as $n \to \infty$ (see Appendix 1). Hence the right-hand side of (10) is positive with probability tending to 1. For finite sample, it can happen that $n^{-1}RSS - \hat{c}^T S\hat{c} < 0$, and we modify the estimator (10) as follows

$$\hat{\sigma}^2 := \max(0, \tilde{\sigma}^2) = \max(0, n^{-1}RSS - \hat{c}^T S\hat{c}). \qquad (14)$$

## 2.3. Asymptotic Normality of the Estimators and Wald Test

Denote

$$\theta = (c^T; \sigma^2)^T, \hat{\theta} = (\hat{c}^T; \hat{\sigma}^2)^T. \qquad (15)$$

The corresponding unbiased estimating vector function is as follows

$$s_\theta = s_\theta(y, w; \theta) = (s_c^T; s_{\sigma^2})^T, \qquad (16)$$

where the components $s_c$ and $s_{\sigma^2}$ are given in (11-12). The estimating equation for $\hat{\theta}$ is the following:

$$\sum_{i=1}^n s_\theta(y_i, w_i; \theta) = 0, \quad \theta \in \mathbf{R}^m \times [0, \infty). \qquad (17)$$

Actually $\hat{\theta}$ is a Borel function of observations $y_1, w_1, \ldots, y_n, w_n$ such that $\hat{\theta}$ is a solution to (17) if (17) has a solution. Under corresponding model assumptions, equation (17) has a solution with probability tending to 1 as $n \to \infty$ (see Appendix 1).

Under a bit stronger model assumption, the estimator $\hat{\theta}$ is consistent (i.e. $\hat{\theta}$ converges to true $\theta$ in probability as $n \to \infty$) and asymptotically normal. Sufficient conditions for the consistency and asymptotic normality of $\hat{\theta}$ are given in Appendix 1. They include finiteness of error moments of some order greater than 4, convergence of sample means of $x$ up to 4th order, and non-singularity of the matrix

$$\mu_2 := \lim_{n \to \infty} \frac{1}{n} \sum_{i=1}^n x_i x_i^T. \qquad (18)$$

The asymptotic covariance matrix $\Sigma_\theta$ can be found from the sandwich formula (see [16]):

$$\Sigma_\theta = A^{-1} B A^{-T}, \qquad (19)$$

$$A := -\lim_{n \to \infty} \frac{1}{n} \sum_{i=1}^n E \frac{\partial s_\theta(y_i, w_i)}{\partial \theta^T}, \quad B := \lim_{n \to \infty} \frac{1}{n} \sum_{i=1}^n E s_\theta(y_i, w_i) s_\theta^T(y_i, w_i). \qquad (20)$$

Here $A^{-T} := (A^{-1})^T$ and the functions are evaluated at the true point $\theta$. We assume that $B$ is nonsingular, and therefore, $\Sigma_\theta$ is nonsingular as well. In our case, $A$ is symmetric, hence $A^{-T} = A^{-1}$ and (19) takes a form

$$\Sigma_\theta = A^{-1} B A^{-1}. \qquad (21)$$

Notice that $\frac{\partial s_c}{\partial \sigma^2} = 0$ and $E \frac{\partial s_{\sigma^2}(y_i, w_i)}{\partial c} = 0$, therefore, the matrix $A$ has a block-diagonal structure

$$A = \begin{pmatrix} \lim_{n \to \infty} E\overline{ww^T} - S & 0 \\ 0 & 1 \end{pmatrix}. \qquad (22)$$

A consistent estimator of $\Sigma_\theta$ can be found as

$$\hat{\Sigma}_\theta := \hat{A}^+ \hat{B} \hat{A}^+, \qquad (23)$$

where $\hat{A}$ and $\hat{B}$ are consistent estimators of $A$ and $B$, respectively. In view of (22) and (20), the matrix estimators can be constructed as follows:

$$\hat{A} = \begin{pmatrix} \overline{ww^T} - S & 0 \\ 0 & 1 \end{pmatrix}, \qquad (24)$$

$$\hat{B} = \frac{1}{n} \sum_{i=1}^{n} s_\theta(y_i, w_i; \hat{\theta}) s_\theta^T(y_i, w_i; \hat{\theta}). \qquad (25)$$

The pseudo-inverse matrix $\hat{A}^+$ has a block-diagonal form

$$\hat{A}^+ = \begin{pmatrix} (\overline{ww^T} - S)^+ & 0 \\ 0 & 1 \end{pmatrix}.$$

The lower right entry $\hat{v}_{\sigma^2}^2$ of $\hat{\Sigma}_\theta$ in (23) is an approximation to the asymptotic variance of the estimator $\hat{\sigma}^2$. Due to the block-diagonal structure of $\hat{A}^+$, it holds

$$\hat{v}_{\sigma^2}^2 = \hat{B}_{\sigma^2} = \frac{1}{n} \sum_{i=1}^{n} s_{\sigma^2}^2(y_i, w_i; \hat{\theta}). \qquad (26)$$

Here $\hat{B}_{\sigma^2}$ is the lower right entry of $\hat{B}$ given in (25). Let $\hat{v}_{\sigma^2}$ be the square root of the approximation (26). Estimated asymptotic standard deviation of $\hat{\sigma}^2$ equals

$$\widehat{se} = \frac{\hat{v}_{\sigma^2}}{\sqrt{n}}. \qquad (27)$$

Let $\sigma_0^2 > 0$ be a given upper bound for the variance $\sigma^2$, that is a known upper estimate of measurement error variance, obtained outside of the model. We test a one-sided compound null hypothesis

$$\boldsymbol{H_0}: \sigma^2 \leq \sigma_0^2 \qquad (28)$$

vs. one-sided compound alternative

$$\boldsymbol{H_1}: \sigma^2 > \sigma_0^2. \qquad (29)$$

Given a confidence level $1 - \alpha \in [0.99; 1)$, according to Wald test we propose the following decision rule:

$$\text{Reject } \boldsymbol{H_0} \text{ if } \hat{\sigma}^2 > \sigma_0^2 + z_\alpha \cdot \widehat{se}, \qquad (30)$$

$$\text{and } \text{do not reject } \boldsymbol{H_0} \text{ if } \hat{\sigma}^2 \leq \sigma_0^2 + z_\alpha \cdot \widehat{se}. \qquad (31)$$

Here $z_\alpha$ is upper quantile of normal law, a critical value of the test. The asymptotic significance level of the test equals $\alpha$, or more precisely, Type I error satisfies relations:

$$\lim_{n \to \infty} P(\text{reject } \boldsymbol{H_0} | \sigma^2 = \sigma_0^2) = \alpha \qquad (32)$$

and

$$\text{for each } \sigma_1^2 \in (0, \sigma_0^2), \lim_{n \to \infty} P(\text{reject} H_0 | \sigma^2 = \sigma_1^2) = 0. \quad (33)$$

The test is consistent, i.e., Type II error satisfies the following:

$$\text{for each } \sigma_1^2 > \sigma_0^2, \quad \lim_{n \to \infty} P(\text{do not reject } H_0 | \sigma^2 = \sigma_1^2) = 0. \quad (34)$$

The rejection criterion (30) can be written in a quite simple practical form. Remember that the residual sum of squares *RSS* was introduced in Section 2.2. For practical use, one can put in (30-31) $z_\alpha = 3$. As a result, we reject $H_0$ with confidence 0.99 if

$$\frac{RSS}{n} > \sigma_0^2 + \hat{c}^T S \hat{c} + \frac{3}{\sqrt{n}} [\frac{1}{n} \sum_{i=1}^n (y_i - \hat{c}^T w_i)^4 - (\frac{RSS}{n})^2]^{1/2}. \quad (35)$$

Otherwise we do not reject $H_0$. We call this procedure *Measurement Error Model Validity test* (MEM-V test).

The p-value for this criterion is calculated as follows:

$$T := \frac{\frac{RSS}{n} - \sigma_0^2 - \hat{c}^T S \hat{c}}{\frac{1}{\sqrt{n}} [\frac{1}{n} \sum_{i=1}^n (y_i - \hat{c}^T w_i)^4 - (\frac{RSS}{n})^2]^{1/2}},$$

$$p = 1 - \Phi(T), \quad (36)$$

where $\Phi$ is cumulative distribution function of standard normal law. Small p-value (usually less than 0.05) indicates strong evidence against the null hypothesis. Large p-value (usually larger than 0.05) indicates strong recommendation not to reject the null hypothesis. In simulation study reported in section 3, under validity of *H₀* the corresponding p-values are large, which leads to the absence of Type I errors (except extreme cases when P/A is close to 1 with very small sample size), and under validity of *H₁* the corresponding p-values are either small or take moderate values larger than 0.05, which leads to certain level of Type II errors.

A disadvantage of the decision rule (30-31) lies in the fact that the test is asymptotic, i.e., it works well for large $n$. But we do not know precisely which $n$ is large enough for typical model parameters. Some empirical suggestions are provided in sections 3 and 4.

Applied to the measurement error model (2-3), the Gleser-Hwang effect (see [11], Section 2.4.1) says that every non-asymptotic confidence set for the regression parameter *c* with positive confidence level will be unbounded with positive probability. For the same reason it is impossible to construct a non-asymptotic test for the above-mentioned null hypothesis, with fixed non-asymptotic significance level.

**Remark 4.** Once $H_0$ is *not rejected*, we infer that the data fit the model (2) - (3) with given upper bound $\sigma_0^2$ for the variance $\sigma^2$. If $H_0$ is *rejected*, then we infer that the term *e* in (2) contains misspecification error and we need moreexplanatory covariates in the linear model (2).

Coming back to the convenient and simple example of Hooke's law in Introduction, suppose that we made measurements of applied force $F_i^{mes} = F_i^{true} + \delta_i$ and measurements of extension $x_i^{mes} = x_i^{true} + e_i$, $i = 1, \dots, n$, where $F_i^{true}$ and $x_i^{true}$ are the corresponding true values. Error variances $\sigma_{\delta 0}^2$ and $\sigma_{e 0}^2$ are recorded in the passport of measuring devices. We assume that $Var\, \delta_i = \sigma_{\delta 0}^2$ is known, while $Var\, e_i =: \sigma_e^2$ is unknown. One can check the null hypothesis $\mathbf{H_0}$: there exists $k > 0$ $with$ $x_i^{true} \equiv k^{-1} F_i^{true}$ and $\sigma_e^2 \leq \sigma_{e0}^2$ .

If $\mathbf{H_0}$ is *not rejected* then we confirm the Hooke's law $F^{true} = kx^{true}$, and if $\mathbf{H_0}$ is *rejected* then we deny the Hooke's law and infer that either extension depends not only on applied force, but on some other covariates, or it depends on force only but in a nonlinear way.

## 3. SIMULATIONS

In order to test the behavior of different statistics associated with the proposed criteria the simulation experiment was performed with different data settings. The following parameters were used to manipulate the data.

1. *Number of covariates* x affecting the outcome y: 2. Both were generated as uniformly distributed random variables but within the different ranges to observe the scaling effect: $x_1$ in a range [0,1.5]; $x_2$– in a range [0,0.3].

2. *Number of observations* in each data set: 30, 100, 500, 5000.

3. Level of *measurement errors* $\boldsymbol{\delta}$ *in x* in (5) was regulated as follows: it was proportional to average level of covariates (0.75 and 0.15 respectively) times a factor 0.01, 0.05, and 0.2. The resulting values were considered as standard deviation for normal random variable, which was added to the actual values of *x*. It means that if relative errors compared to average were 1, 5, and 20%, the individual relative errors in each observation could be much smaller or larger.

4. A square root of variance $\sigma^2$ in (28), *a measurement error for y*, was set, similarly to ones for *x*, as 1%, 5%, 20% of the *y* average value.

5. *"Presumed/Actual" (or P/A) ratio* $\sigma_0^2/\sigma^2$ was set at three levels: 0.5, 1, and 2. For each level P/A ratios were generated as random within a narrow range; as a result they were located from 0.4 to 2.4 range, i.e. assumed values could be 2.5 times smaller or 2.5 larger than the actual generated errors.

6. Variance $\sigma_0^2$ in (28) is a researcher's *guess about the measurement error variance* $\sigma^2$ of the measurement error in dependent variable. It was generated as a product of the error $\sigma_0^2$ and the P/A ratio. It reflects a realistic situation that a researcher should be more or less aware about the real measurement level and do not apply too high or too low P/A ratio to make an estimate.

7. Models of two types were constructed: when *y* was a linear function of the two *x* covariates with errors, as described above, and when *y* was also containing the third covariate $x_3$ (representing all unobserved factors), which created another source of *y* variance, different by its nature from the *y* measurement error. They were named *models without or with inclusion*, respectively. The $x_3$ was generated as uniformly distributed variable in range [0, k], where k = average $(x_1, x_2)$*f, f takes values 0.1; 1; 3. If f=0.1 – the added covariate should not affect *y* too much, while with f=3 it is a very significant contributor into *y* variation. All estimates of the parameters were made, of courses, without accounting for this covariate, because in practice nobody knows about its existence and character.

Combination of the listed conducting parameters – inclusion of parameter 2 (a sample size from the list above); measurement levels (parameter 3) for $x_1$, $x_2$, *y* and P/A ratio (parameter 5) – yields 162 types of data. For each 30 random datasets were generated, what gives 4860 data sets for each given data size. It allows to calculate the errors of the first and second types across all parameters' combinations with confidence probability 0.99 (32, 34) and p-value (36), along with other characteristics.

Some simulations results are summarized in table 1. In its first section the table shows the average values across all respective datasets. It reduces volatility of the indicators, for it will be much higher if consider more specific groups. Still, some regularity is to be seen.

*Percent of Ho rejections* is visibly increasing when sample size is increasing - with all other conditions are the same, from 13% for 30 to 52% for 5,000 cases. Especially sharp the growth is for situations when Ho is false, from 0 to 86%. This effect is not present only when unobservable covariate is not added and Ho is true – in this case test always works correctly (not rejecting hypothesis), with just 1% of errors for 30 observations, what expected.

Ideally, a rejection rate should not depend on *n:* it has to reflect just specific features of the data, like interplay between measurement errors, as it was intended for. But dependence happens; it sheds light not only on this specific problem, but on any procedure of the statistical estimation. Typically, since a sample size *n* is in a denominator of the formulas for sampling errors and similar statistics, the larger sample – the higher probability that certain types of "significant differences" will be always found, with any confidence level. This fact is one of the key issues of the mentioned above "p-value crisis" [13, 14] and, particularly, inspired the proposal of the so called *d-values*, which is free of this distortion effect [18] (in fact it just shifts the difficulties and responsibilities from statistician to the decision maker, providing her with probability, not with hard cut threshold as p-value with an attached confidence does.) This problem is lurking in our case as well: the larger a sample, the more likely that Ho will be rejected, no matter what. Perhaps, to smooth it one should use probabilities instead of p-values in a sense of [18], but issue remains controversial. On the one hand, the larger – the better and we incline to believe more in estimate based on 5,000 rather than 30 cases. On the other hand – mechanical raise of rejections percent with sample size growth seems unnatural and disturbing. Maybe, one could just use caution looking at the rates of growth in table 1 and coordinating his/her results with it (for example, use bootstrap for double check and so on). Nevertheless, we continue, with all those remarks, in traditional fashion, leaving controversies aside.

*Coefficient of determination* $R^2$ shows level of *y* approximation by *x* covariates in each of specific dataset. It is, generally, doesn't depend on sample size and other circumstances, but is strongly affected by adding the hidden covariate into the model (see table 2).

The table 1 does not show the *type 1 error* for simple reason: it is always equal to 0. Even model with sample size 10 (not shown here) yields the same output. This is quite amazing result, which talks a lot about selectivity and directionality of the criterion. The only situation when *type I appears* is when presumed variance is practically equal to the actual variance (i.e. when P/A ratio is around 1) and sample size is small, like 30 or less. This is quite paradoxical: if one knows exactly what errors in *y* variable are, the probability to make a mistake in model's specification raises! But, at the second glance, the reason for that is clear: the closer two values, the higher the chance, that small random fluctuation shifts the decision into the wrong side. Relation of the P/A ratio and error of the type II is also non-trivial.

*Errors of type II* behave in predictable way: their average level for all situations drops from very high 71% for small samples size 30 to more acceptable level of 19% for 5,000. Even sharper decrease observed for the situation, when $H_0$ is false, and yet the unobserved covariate is not in a model, from 99% to 14%, i.e. if one wants to make sure that measurement errors (the only source of variance here) do not disturb model too much – only big number of observations should be available to conclude it with good confidence; even samples as big as 500 observations are not enough for that. It's a very important conclusion for practical purposes (see [4]). On the other hand, when the latent covariate is there – even 5,000 cases yield the high errors, 30%. However, the real influence of this hidden covariate depends very much on its comparative size to the observed covariates (see below).

Table 1. Average values and correlations of p-values for different indicators[1]

| Indicators | Inclusion of $x_3$ | Sample size | | | |
|---|---|---|---|---|---|
| | | 30 | 100 | 500 | 5000 |
| EIV test, % of rejection | All cases | 13% | 27% | 41% | 52% |
| | Yes | 25% | 43% | 52% | 60% |
| | No, Ho true | 1% | 0% | 0% | 0% |
| | No, Ho false | 0% | 21% | 57% | 86% |
| Determination, $R^2$ | All cases | 73% | 73% | 73% | 72% |
| | Yes | 63% | 62% | 63% | 62% |
| | No, Ho true | 82% | 83% | 82% | 82% |
| | No, Ho false | 82% | 83% | 82% | 82% |
| Error of type II | All cases | 60% | 45% | 30% | 19% |
| | Yes | 71% | 51% | 39% | 30% |
| | 0, Ho true | n/a | n/a | n/a | n/a |
| | 0, Ho false | 99% | 79% | 41% | 14% |
| p-value | All cases | 0.47 | 0.41 | 0.38 | 0.37 |
| | Yes | 0.34 | 0.29 | 0.27 | 0.26 |
| | No, Ho true | 0.89 | 0.90 | 0.94 | 0.97 |
| | No, Ho false | 0.33 | 0.19 | 0.09 | 0.02 |
| **Correlations with p-values** | | | | | |
| EIV test, % | All cases | -0.47 | -0.61 | -0.77 | -0.91 |
| | Yes | n/a | -0.69 | -0.79 | -0.93 |
| | No, Ho true | 0.00 | 0.00 | 0.00 | 0.00 |
| | No, Ho false | -0.12 | -0.36 | -0.56 | -0.57 |
| Determination, $R^2$ | All cases | 0.43 | 0.36 | 0.30 | 0.27 |
| | Yes | 0.04 | 0.43 | 0.37 | 0.34 |
| | No, Ho true | 0.04 | -0.05 | -0.14 | -0.19 |
| | No, Ho false | 0.23 | 0.10 | 0.03 | -0.02 |
| Error of type II | All cases | 0.54 | 0.69 | 0.79 | 0.93 |
| | Yes | -0.02 | 0.03 | 0.20 | 0.37 |
| | No, Ho true | 0.00 | 0.00 | 0.00 | 0.00 |
| | No, Ho false | 0.12 | 0.36 | 0.56 | 0.57 |

[1]n/a stands for "not available", either theoretically (as in a row for type II errors) or actually, when all tests showed one (positive) result and correlation was impossible to calculate

As for p-values, the most interesting fact is this: if hidden covariate was not there and Ho is hold, they are much larger than ones for situation where $H_0$ is false – about 0.9 vs. ones from 0.33 to 0.02. This is understandable, because in the case of true $H_0$, huge p-values indicate our confidence in accepting the null hypothesis, and in case of false $H_0$, small p-values indicate our confidence in rejecting the null hypothesis. To look at p-values closer, the lower part of the table 1 presents their correlations with other statistics in original data (each correlation was calculated for 4860 observations for each data size or smaller, depends on number of filtering conditions.)

Correlations of p-values under false $H_0$ are strongly negative with growth of the *sample size*: the larger sample, the less p-value, what follows from considerations discussed above.

*Determination* is not strongly related with p-value in scenario of the correct model (without adding covariate), but positively correlated with p-value when the latent covariate is added.

*Errors of type II* are associated with p-values differently for correct and misspecified models, but for practical purposes it is important to look at the correlation "for all" when both cases are combined. The reason is –no one usually knows whether a model is specified correctly or not. So, one may expect to have smaller p-values with errors if sample size is small and larger p-values for the large sample size (the correlations go from negative to positive). In other words, for large sample size it will be harder to make an accurate decision about correctness of the model. But interestingly enough, if the model is specified correctly and Ho is indeed true (i.e., measurement errors are not too large) – the correlation is zero for any sample size, the only situation where bias of the large sample size for p-values, discussed above, is avoided. This means that in this case we very confidently accept the null hypothesis, whatever sample size is.

Table 2 shows some indicators under different angle. Interestingly, different levels of errors either in *y* or in *x* do not affect strongly any statistics. But effect of $x_3$ size is very strong for all of them: with raising relative importance of added covariate percent of test rejections grows from 20 to 84; determination drops from 0.82 to 0.27; errors of type II sharply reduced from 77% to 5%, and p-values – from 51 to 0 percent (!). So, for the hidden covariate to be the main factor affecting everything the size of the effect is a key, not the very fact of it. If one compares the statistics from the first row, when the hidden covariate represents just about 10% of the most important correct covariate, and compares that with the last row, when it is 3 times larger than that – the difference is startling and conclusions are completely different. Alas, we usually do not know not only the size of the hidden covariate, but the fact of its presence.

Effect of the hidden covariate itself depends very much on the size of the sample; fig.1 provides the illustration of that for p-values. If for non-influential hidden covariate small p-values are not obtainable even with huge dataset, for very important half of those is below the common threshold 0.05 even when sample size is just 100. Conversely, for unimportant covariate one will never have the confidence in a model, while for important one the share of the high p-values doesn't exceed 50% even for sample size 30 and drops down to 28% with 5,000.

In total, out of all 216 (54*4) combinations of the conducting parameters, size included, small average p-values (less than 0.05) were observed in 40 cases (19%). They all appear *only* when $x_3$ is added, with 6 (15%) when the $x_3$ is about the same size as actual ones and the rest – when it exceeds others in 3 times. Thus, the reliable separation (with small p-value) of the correct and incorrect models is practically possible only if some unknown and influential covariates are present; if not – the decision will not be strongly supported, even when level of errors in variables is as big as 20%, as in our experiments. It, however, does not mean that if the third covariate is added the correct model always will be detected with high power: in 60% of all cases with inclusion (out of which 45% have $x_3$ of about the size of the correct covariates $x_1$ and $x_2$) p-value is very high. These observations emphases the fact, how difficult is to make a correct decision about the truthful models.

Table 2. Average values of indicators as functions of scaling parameters

|  |  | Test rejections,% | Determination, $R^2$ | Error of type II | p-value |
|---|---|---|---|---|---|
| Y measrement error, $\sigma$ | 0.01 | 0.26 | 0.83 | 0.37 | 0.37 |
|  | 0.05 | 0.37 | 0.76 | 0.39 | 0.42 |
|  | 0.20 | 0.36 | 0.59 | 0.40 | 0.44 |
| X measurement errors, $\delta$ | 0.01 | 0.29 | 0.75 | 0.41 | 0.38 |
|  | 0.05 | 0.39 | 0.74 | 0.33 | 0.41 |
|  | 0.20 | 0.31 | 0.68 | 0.41 | 0.44 |
| Factor f for X3 | 0.10 | 0.20 | 0.82 | 0.77 | 0.51 |
|  | 1.00 | 0.31 | 0.79 | 0.61 | 0.35 |
|  | 3.00 | 0.84 | 0.27 | 0.05 | 0.00 |

Another, rather surprising, aspect of the estimation procedure is revealed when one considers how confidence changing when quality of our guess about the real variance for *y* is approaching the true level. Fig. 2 shows how p-values are changing when P/A ratio increases. This chart represents actual values, as in table 2, not the average ones. If $\sigma_0^2$ is much smaller then $\sigma^2$ (left part of the P/A ratio axis) – the decision in favor of the alternative hypothesis are made without errors, all p-values are close or equal to zero; when it is much larger – the decision in favor of the null hypothesis are made without errors, all p-values are close or equal to one. And only in narrow area surrounding the point where P/A ratio equals one, p-value is increasing very fast with visible sharp slope.

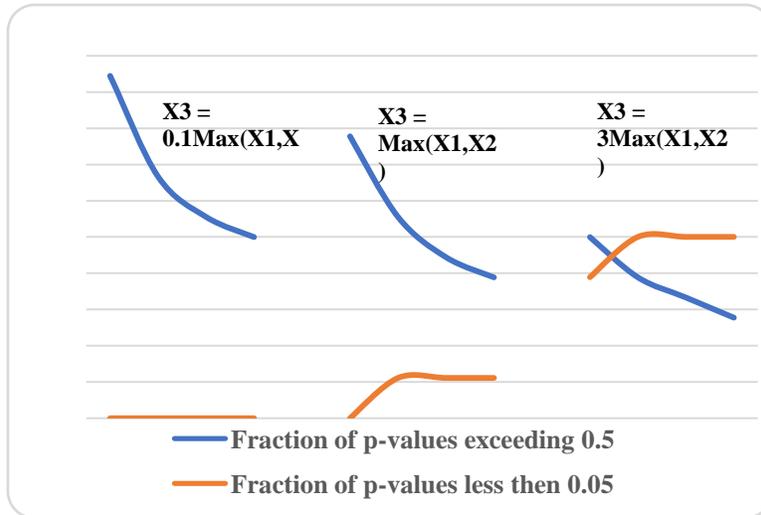

Fig.1. P-values for models with hidden covariates $x_3$

Thus, it makes sense to presume either small variance or big, if one wants to make accurate estimation of the model. If the presumed variance $\sigma_0^2$ is quite close to the true one, then it is difficult to make a reliable decision which of the two values is larger. If $\sigma_0^2$ is rather far from $\sigma^2$, then our decision is precise, but the accepted hypothesis is rather weak, i.e., we decide only that the true variance is larger than some small quantity or decide that $\sigma^2$ is smaller than some large quantity.

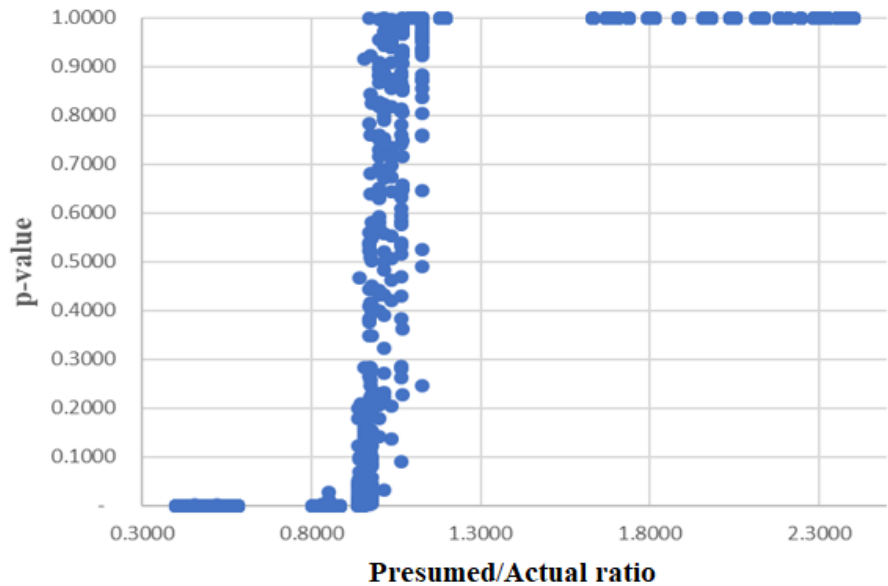

Fig.2. P-value and Presumed/Actual ratio; no Inclusion of the third covariate; 5,000 cases in each dataset

In practical case, where the covariance matrix of measurement errors in $x$ is unknown, we can use the test with presumed value $S = \kappa \hat{S}_w, \kappa \in [0,1)$, as proposed in Appendix 2. Here k is a share of error variance in the total x variance under assumption that relative errors are the same for each x variable. In Fig. 3 we see that for $\kappa \leq 0.54$, the corresponding p-value is close to 1, and for $\kappa \geq 0.62$, the p-value is close to 0. With certain precision, this corresponds to the intervals where the asymptotic value $A_\infty$ of the test statistic is negative ($\kappa < 0.62$) and positive ($\kappa > 0.62$), respectively. See Appendix 2 for the relation between p-values and the behavior of $A_\infty$.

In fact, from all which was observed from simulations so far one main idea could be informally derived: the proposed criterion somehow separates situations with and without "improper covariates". It behaves very differently in those two scenarios, what was clear from the tables and charts above. Could we use it for this purpose directly? If yes, it would be very attractive idea. If one may confidently say that the given model with all assumptions about measurement errors contains or not the unobserved covariates, it will be an important result in statistical modeling. This type of criterion should, ideally, replace many heuristic ones in a large area of research for the "best variables selection" (see, for example, [19] and references therein).

The proposed criterion does not answer this question strictly, but leads into right direction. It turned out, that the key results from [20] combined with ones from this paper could be used to formulate the correct criterion for answering the question: *are there hidden independent covariates in data or not*. But this is a special topic to be considered in a separate paper.

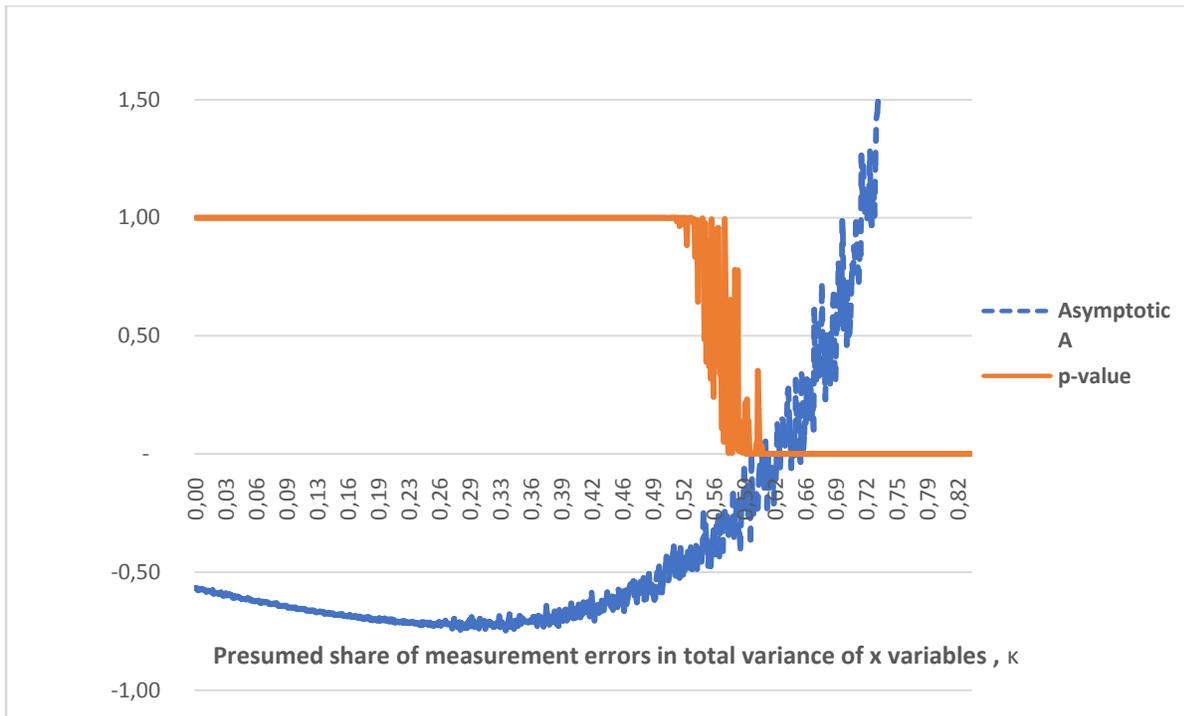

Fig.3. Asymptotic value of the numerator in (A13) of the test statistic and
p-value of the test as a function of $\kappa$

## 4. REAL DATA EXAMPLE: CALORIES IN FOOD

To demonstrate how the proposed test may work in practice let's consider a simple example with real data, which were discussed at length in [5] and especially [4] in connection with casual modeling problem. It is a collection of 872 types of foods, were the content of *fats, proteins, carbohydrates* and total amount of *calories* is measured, all per 100 grams of foodhttps://www.csun.edu/science/ref/spreadsheets/xls/nutrition.xls. Since calories could be obtained only from those three ingredients (not from, say, water or fiber), the model with three independent covariates is *causal, linear* and *complete*. The remaining variance in *y* (calories), if any, could be explained only by measurement errors in *x* and *y*. In that aspect a model is ideal for the study, for not that many real cases of such a nature may be easily found; it exactly fits description of our model in (2) and (3). Let consider application of the proposed test to two situations, as described earlier: for complete model and for the model with added hidden covariate. We start with the simplest case.

**Complete model: no hidden covariates.** Certain number of foods does not have one of the ingredients. We took the subset without fat (241 foods), i.e. just with two independent covariates, *Carbs* and *Proteins* (to be called $w_1$ and $w_2$ in (3)). As expected, the model shows very high goodness-of-fit, with OLS estimates of the coefficients 3.91 and 3.94 and with $R^2$=95.6%. Those coefficients match perfectly those widely known in nutrition literature, reported usually just as equal to each other, 4.

Now, let's check the hypothesis about validity of this model. As formulated in (28-35), the test compares two errors in *y*, measured as variances, the true one and the estimated. Let say, we have no idea how big or small the real error in *y* is, as well as we do not know how big are the errors in *x* covariates (which is indeed the case – the publishers of the data did not say anything about errors). In order to test the validity, one has to make those assumptions in kind of plausible way and make test's calculations. But what is plausible?

The best way to judge it is to use *relative errors*, i.e. fraction of the presumed **δ** to the standard deviation of the respected covariate *w* or response variable *y*. This fraction cannot be higher than 1, and we run the estimates in this whole range of possibilities. Another assumption was that relative errors for each covariate are the same (it could be easily relaxed without loss of generality). P/A ratio was also set in a respective range in such a way, that its maximal value does not allow the measurement error exceed the standard deviation of *y*. The results could be briefly summarized as follows.

1. When critical value of the test in (30) is set as 3 (which yields the confidence probability P=0.99), as typically people do, the Ho was not rejected under any conditions.

2. When it was set to 2 (P=0.95), the rejection took place only if presumed measurement errors, both for *x* and *y*, are very small - see fig.4. Two curves are practically not distinguished from each other; if presumed error for *y* reaches, say, 10% - no rejection will take place. The bar on a chart shows the area of errors in x when Ho was rejected. The maximal p-value in this area was around 0.02, i.e. very small. The further increase of the errors in *x*, as seen, doesn't affect the rejection rate but increases p-value.

3. The S-shape type of the curve for each presumed level of the relative error for *y* resembles one from fig.2. The *rule of thumb recommendation* may be provided, that the presumed variance for *y* should be about RSS/(*n-m*), where *n* is number of observations and *m* - number of covariates. In our case it gives relative error close to 90%, which is in concordance with the earlier observation that larger presumed values are more preferable than the smaller one for test to work more confident. Yet large values, as fig. 4 shows, would definitely eliminate any probability of rejection.

4. Those conclusions are independently supported by considerations related with sign of the numerator in (A14). Despite the fact that sample size here is not that big, a chart like one on fig. 3 shows the similar non-monotone curve for $A_\infty$ (fig. 5). It crosses zero at about 10% point, which corresponds with beginning of p-value raise on fig.4.

5. The described effects are intuitively clear, even if not immediately obvious. Imagine, we presume that data have no measurement errors at all, the function is strictly linear. Of course, Ho should be rejected, because in fact we do observe unexplained variance in *y* variable! The test demonstrates just that. In this specific dataset, the plausible level of errors in *y* is very high (see above the rule of thumb), and even for that reason we should not reject hypotheses. But more probably, the real error of measurement in food ingredients is much higher that 5% on a chart – which also tells us that Ho should not be rejected. And if we want to be more confident (use 0.99 probability) – then it is even more true.

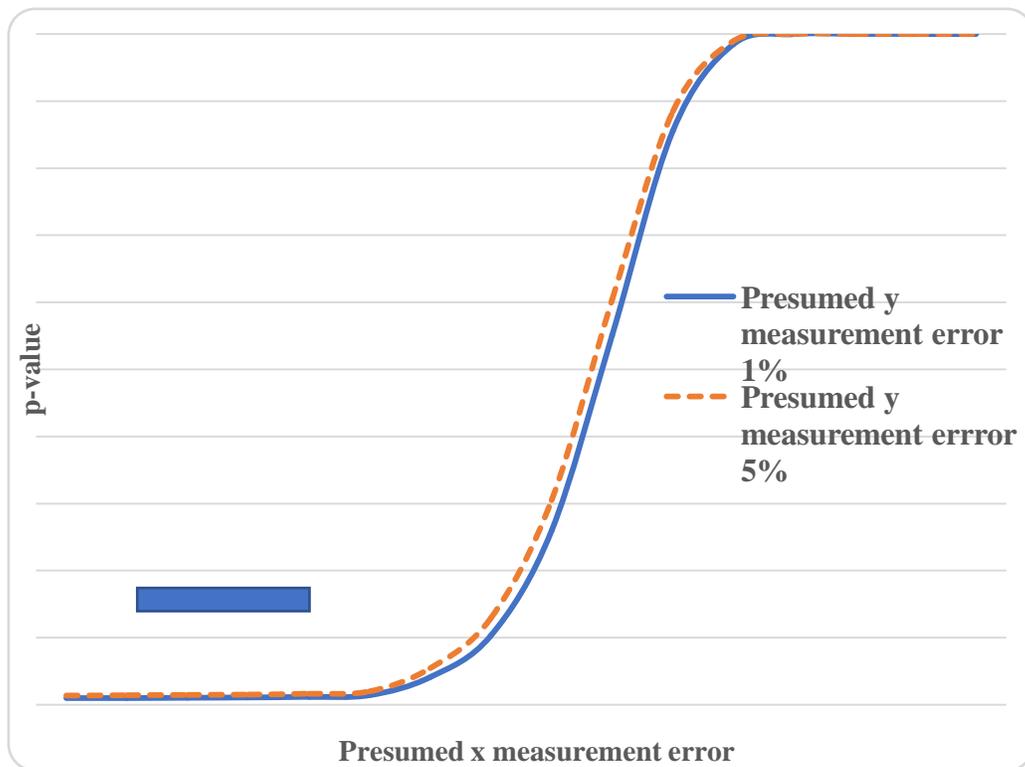

Fig.4. MEM-V test for complete model *Carb* and *Protein*; $R^2=0.96$
(bar shows the area when Ho was rejected with confidence probability P=0.95)

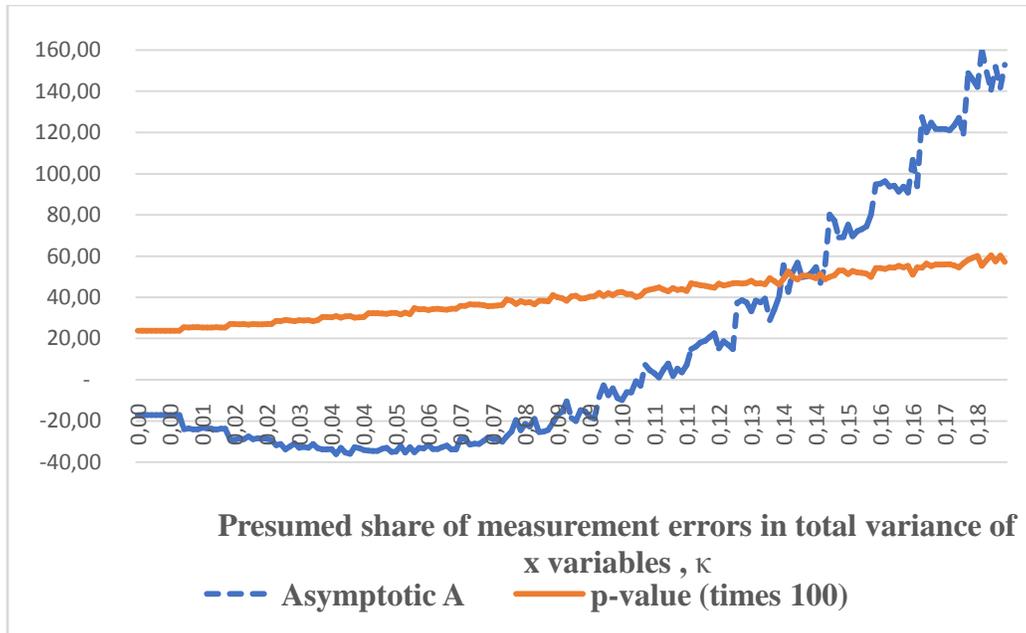

Fig.5 Asymptotic value of the numerator in (A13) of the test statistic and p-value of the test as a function of κ for *Carb* and *Protein* model

So, overall, the MEM-V test provides very strong evidence that the model is valid, which is indeed true. What if it is not?

**Misspecified model: hidden covariate.** New let's consider much more common scenario, when there is a hidden covariate(s) in a model, but researcher has no idea about it. For testing we may use the entire food dataset, but include in a model the same two covariates as before assuming that the very notion of *Fat* is unknown to the curious and slim statistician. Expectedly, the quality of the regression model drastically dropped, with $R^2$ just 15.3% (vs. 95.6% for the complete case earlier), and coefficients severely biased: 2.5 and 5.5 vs. correct 4 for both covariates. This model, however, perfectly satisfies the standard statistical requirements: t statistics for coefficients are 10.4 and 8.3, much higher than usually required 3, and p-values are practically zero, thousand times smaller than the infamous 0.05, which once again tells about fragility of all that type of "requirements". Can the proposed test conclude that something is still wrong here?

Yes, it can. Fig. 6 shows how the four statistics are constructed for different assumptions about errors in covariates and dependent variable. The shape of each curve is somehow similar to S-like curve on fig.4 (if one zooms closer to inflection point), but most importantly – the hypothesis that model is correct is rejected immediately and on entire range of assumptions about errors, what follows from the horizontal bars of rejections. Difference in presumed errors for *y* practically is negligible. The test immediately detects, that despite all "perfect" statistics, the model is invalid. Note, it does it with very high confidence, when P=0.99.

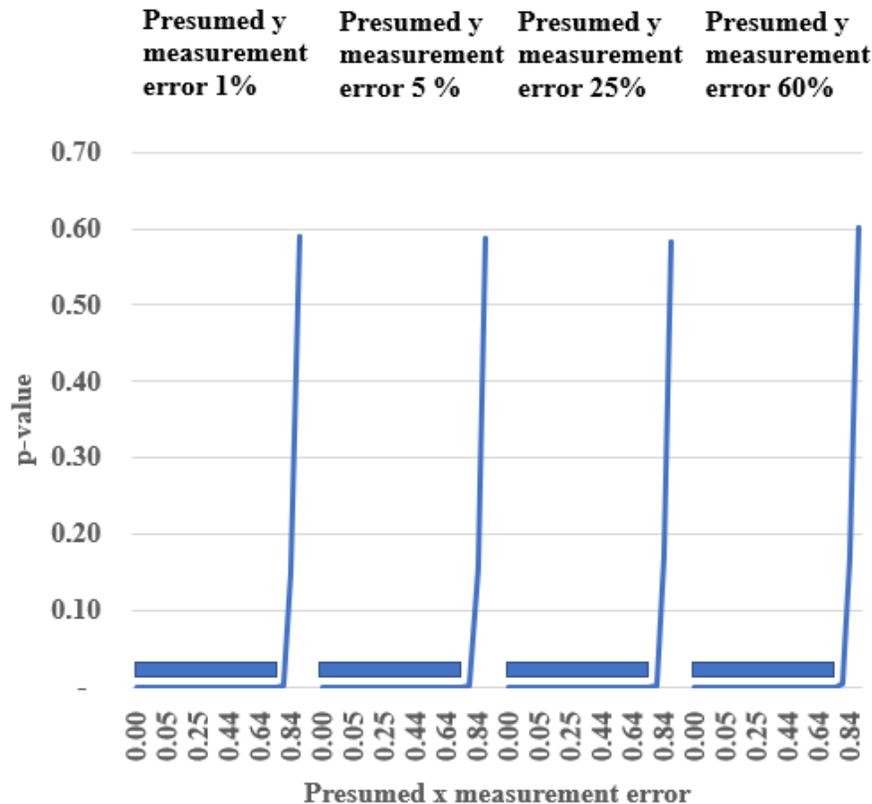

Fig.6. MEM-V test for incomplete model with *Carb* and *Protein*, hidden *Fat*; $R^2=0.15$
(P=0.99, horizontal bars show the values when Ho was rejected with p-value about 0)

But, possibly, a test worked very well because the original model was too weak, judging by common sense about low determination (15%), not by mysterious other indicators (t-stat and p-value), stating just the opposite? Let's check it on another pair of covariates, giving much more formal "confidence" that model is good.

It turned out, that if we exclude *Protein*, but leave *Carb* and *Fat*, the model will seem just perfect: the determination is 95.4% (!), coefficients are very close to the real ones – 3.6 and 8.8 (vs. 4 and 9), and, of course, t statistics (63 and 129) and p-values (0) are much better any accepted required levels. But the model is incomplete, and we know that. What about the test? It doesn't disappoint even in this difficult for detection situation. The results are shown on fig. 7.

In the same fashion as before, the test rejects the hypotheses, but, remarkably, does it for much shorter range of the available presumed errors for covariates. This range becomes even shorter with increasing of errors for *y*: up to about 16% for small errors in y and up to 10% when error for *y* is around 60%. It makes a lot of sense. When determination is very high and the model is valid, big measurement errors in covariates are actually impossible, there is no room for them, and Ho almost never is to be rejected or rejected not with high confidence on very narrow range for errors (like in fig. 4). But when the same high determination takes place on invalid model, as here, the restriction for errors in covariates remains, but its range becomes much larger (in three times) and confidence in rejection much higher (0.99 vs. 0.95). Thus, even in this almost extreme case test worked out and detected the invalidity, which would escape attention otherwise. The decision is justified with less confidence, in general, than it was on fig.6, with determination 6 times smaller, but still justified.

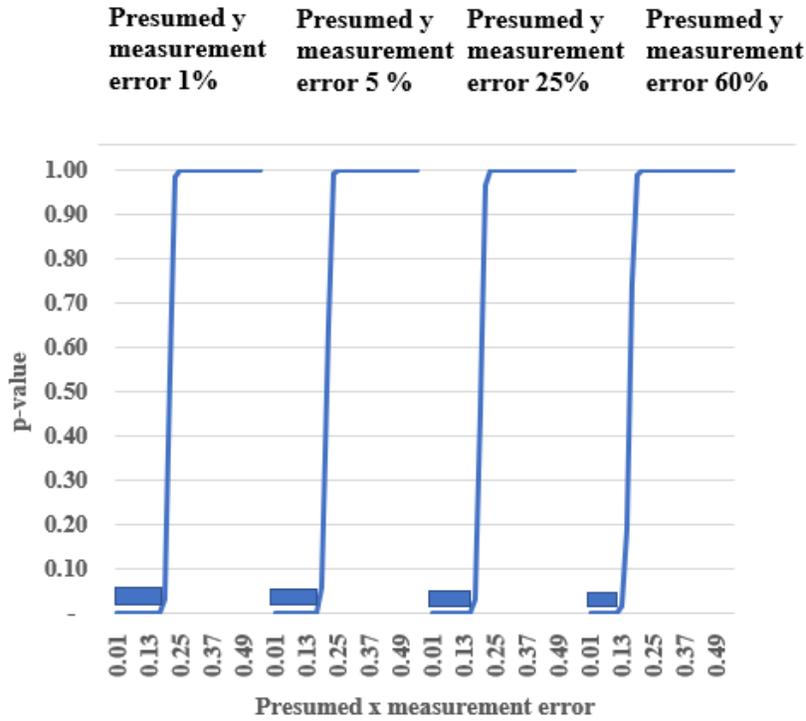

Fig.7. MEM-V test for incomplete model with *Fat* and *Carb*, hidden *Protein*; $R^2=0.95$
(P=0.99; horizontal bars show the values when Ho was rejected with p-value about 0)

Those examples show how the MEM-V test could be used in practice. The radical difference between superficially almost identical models described in fig. 4 and fig. 7 shows it very well: in both cases approximation is extremely high ($R^2=0.95$), but test convincingly reveals the qualitative difference. It seems it can play an important role in separation of the complete and incomplete models in the very common situation of errors in variables for any linear models.

In summary, here are short recommendations how to use the proposed measurement errors model validity test in practice.

1. If error variance $\sigma_e^2$ of *y* variable and error covariance matrix $S_\delta$ of *x* variable are known (like when measurement devises with given calibration are used for both *y* and *x*), use the test directly with $S = S_\delta$ and $\sigma_0^2 = \sigma_e^2$.

2. If error variance $\sigma_e^2$ of *y* variable is known but error covariance matrix of *x* variable is unknown, then set $\sigma_0^2 = \sigma_e^2$ and $S = \kappa \hat{S}_w$, $\kappa \in [0, 1)$, where $\hat{S}_w$ is a sample covariance matrix of *w*(see Appendix 2), and use the test for different presumed values $\kappa$. Plot the dependence of p-value of the test as a function of $\kappa$, i.e. build a chart like the one on fig. 3. Set the confidence level (e.g. $P = 0.99$). Indicate the intervals for $\kappa$ where the corresponding p-value is less than $1 - P$ (where $H_0$ is rejected and the model is invalid) and the intervals where corresponding p-value is greater than or equal to $1 - P$ (where $H_0$ is not rejected and the model could be valid). Then select one or another interval for $\kappa$ and make the corresponding decision according to your ideas about possible level of measurement errors of x variable. Since $\kappa$ represents the error's share in total x variance, if $\kappa$ is less than presumed level of errors – the Ho is accepted. Keep in mind though, that it is not a universal situation: sometimes a plot for A8 may have several crossing points with horizontal axe. If you are not able to make the final choice, then you need more information about measurement errors of *x* variable.

3. In a case when error variance $\sigma_e^2$ of *y* variable is known and the error covariance matrix $S_{(1)}$ of some components $x_{(1)}$ of *x* variable is known, while for the rest components $x_{(2)}$ it is unknown, make the analysis similar to point 2 above with a block diagonal matrix *S*, with blocks $S_{(1)}$ and $\kappa \hat{S}_{W_{(2)}}$, $\kappa \in [0, 1)$, where $\hat{S}_{W_{(2)}}$ is a sample covariance matrix of $w_{(2)}$, surrogate data for unobservable components $x_{(2)}$.

4. If error variance of *y* variable is unknown and error covariance matrix $S_\delta$ of *x* variables is known, then the proposed test is not applicable. But another test developed in [20] can be used. We intend to study that test in a future research.

5. If error variance of *y* variable is unknown and error covariance matrix of *x* variable is unknown or only partially known, one can try the test from [20] with matrix *S* selected as in points 2 or 3 above. This case will be also studied in future research.

6. Finally, suppose that error variance of *y* variable $\sigma_e^2 = \lambda \sigma_{nom}^2$ and error covariance matrix of *x* variable $S_\delta = \lambda S_{nom}$ with known characteristics $\sigma_{nom}^2$ and $S_{nom}$ (those nominal values are usually taken from passports of measurement devices) and unknown positive factor $\lambda$. It may happen, when, for example, all variables are measured with the same device, but accuracy of this device is not precisely known, i.e. we do not know the exact $\lambda$. Under a presumed upper bound $\lambda_0$, one can check the null hypothesis: "$\lambda \leq \lambda_0$" (this will be studied in a future research); and without a presumed upper bound, one can use a test developed in [22].

## CONCLUSION

We considered the classical measurement error model (2-3), where the variance-covariance matrix *S* of the measurement error $\delta$ is known (this can happen, e.g., in situation of repeated observations of covariates or when we trust the passport information about the precision of measurements) and the variance $\sigma^2$ of the error in response is unknown. Given upper estimate for $\sigma^2$, we checked a hypothesis that the true value of $\sigma^2$ does not exceed the upper estimate. In case the hypothesis is rejected, we infer that either error in response variable contains misspecification component (i.e., model lacks explaining covariates), or actual level of measurement errors in data exceeds the upper estimate. Further study to be published will demonstrate that this mixture can also be untangled: situation with misspecification could be separated from the one with measurement errors only.

Theoretical exploration, intensive numerical simulations and practical example allow listing the following important features of the proposed MEM-V test:

- under almost all circumstances it separates valid and invalid models with error of type I equal to zero, but with very different level of errors of type II;

- it is very sensitive to the presence of the hidden (unobservable) covariate and detects it even in situations, when everything tells the opposite: a model seems close to perfect (very high approximation, very good values of traditional indicators, like p-values or t statistics, etc.);
- sample size $n$ plays generally an expected role in the test – the level of the errors of second type is dramatically decreased when $n$ raises, yet it (not always) also demonstrates the known problem, when tendency to reject hypotheses is increasing with rise of the sample size;
- to make a confident conclusion that only errors determine $y$ variance, while hidden covariates are not in a model, one should have many cases available (more than 500 at least);
- the most important factor among others (like errors in variables, assumptions about the error level, etc.) affecting the correct decision is the relative size of the hidden covariate: when it is high compared to observed ones, the quality of the decision is dramatically improved (with p-value close to zero), although the test works also when it is not that high;
- dependence of the correct decision on our assumption about errors in independent variable is, counter-intuitively, very low, or even close to zero, much smaller than the same assumptions about errors in covariates;
- We did not consider the case where $\sigma^2$ is unknown and $S = \lambda S_0$ with known matrix $S_0$ and unknown scalar factor $\lambda$, because in this case the model (2-3) is not identifiable and regression parameter $c$ cannot be estimated consistently (this is shown in [11] for the structural normal model, where errors and regressor $x$ are normally distributed). In future we intend to construct Wald test in the important case, where the variance-covariance matrix of the augmented error $(e, \delta^T)^T$ is known up to a scalar factor, while some components of the augmented error are allowed to be identical zero.

A survey of more than 1,500 scientists in different fields revealed that more than 80% of the participants consider as a most important improvement to be made to increase the reproducibility of the scientific results is "better understanding statistics" [14]. Problem of measurement errors, with all its importance and universality, is still out of the mainstream statistical thinking. But without "better understanding it" the lack of reproducibility will embarrass scientific community as it did before and does now. Our study shows a possible way to address some critical topics in that undeservedly neglected area.

**Acknowledgement.** Authors are grateful to Prof. Y. N. Tyrin for inspirational discussions and suggestions during last several decades.

# APPENDIX 1: ASYMPTOTIC PROPERTIES OF ESTIMATORS

Introduce additional conditions on the model (2-3) described in Section 1. We start with moment conditions on error terms.

(iii) For some fixed $\tau > 4$, $E|e|^\tau < \infty$ and $E||\delta||^\tau < \infty$.

Next come conditions on empirical moments of regressor. Below $x(i)$ denotes $i$th coordinate of covariate $x$. Remember that bar means averaging over $i = 1, \dots, n$.

(iv) There exists $lim_{n \to \infty} \overline{xx^T} =: \mu_2$, and $\mu_2$ is a nonsingular matrix.

(v) There exists $lim_{n \to \infty} \bar{x} = \mu_1$, and for each $1 \leq i \leq j \leq k \leq l \leq m$, there exist
$$lim_{n \to \infty} \overline{x(i)x(j)x(k)} =: \mu_3(i,j,k), lim_{n \to \infty} \overline{x(i)x(j)x(k)x(l)} =: \mu_4(i,j,k,l).$$

(vi) There exists $C > 0$ such that for all $n \geq 1$ it holds $\overline{||x||^\tau} \leq C$, where $\tau > 4$ comes from condition (i).

Final condition involves the estimating function $s_\theta(y, w; \theta)$ introduced in (16), (11), (12). The existence of the limit below will follow from conditions (i-v).

(vii) At the true point of parameter $\theta$, the matrix

$$B := \lim_{n\to\infty} \frac{1}{n}\sum_{i=1}^n E s_\theta(y_i, w_i; \theta) s_\theta^T(y_i, w_i; \theta) \tag{A1}$$

is nonsingular.

Consider also weaker conditions.

(viii) At the true point of parameter $\theta$, the matrix

$$B_{\sigma^2} := \lim_{n\to\infty} \frac{1}{n}\sum_{i=1}^n E s_{\sigma^2}(y_i, w_i; \theta) s_{\sigma^2}^T(y_i, w_i; \theta) \tag{A2}$$

is nonsingular.

(ix) At the true point of parameter $\theta$, the matrix

$$B_c := \lim_{n\to\infty} \frac{1}{n}\sum_{i=1}^n E s_c(y_i, w_i; \theta) s_c^T(y_i, w_i; \theta) \tag{A3}$$

is nonsingular.

Below $\to^d$ denotes convergence in distribution.

**Theorem.** *For the model* (2-3), *assume conditions (i-vi). Then the next statements hold true.*

1. The estimating equation (17) has a solution with probability tending to 1 as $n \to \infty$, and the resulting estimator $\hat{\theta}$, which is defined in (15), (8), (14), converges in probability to the true value $\theta$, moreover

$$\sqrt{n}(\hat{\theta} - \theta) \to^d N(0, \Sigma_\theta), \tag{A4}$$

with

$$\Sigma_\theta = A^{-1} B A^{-1}, \tag{A5}$$

$$A = -\lim_{n\to\infty} \frac{1}{n}\sum_{i=1}^n E \frac{\partial s_\theta(y_i, w_i)}{\partial \theta^T}, \quad B = \lim_{n\to\infty} \frac{1}{n}\sum_{i=1}^n E s_\theta(y_i, w_i) s_\theta^T(y_i, w_i). \tag{A6}$$

2. Let $\hat{A}$ be defined in (24) and

$$\hat{B} = \frac{1}{n}\sum_{i=1}^n s_\theta(y_i, w_i; \hat{\theta}) s_\theta^T(y_i, w_i; \hat{\theta}). \tag{A7}$$

Then $\hat{A}$ and $\hat{B}$ converge in probability to $A$ and $B$, respectively, and $\hat{\Sigma}_\theta = \hat{A}^+ \hat{B} \hat{A}^+$ converges in probability to $\Sigma_\theta$.

3. Under additional condition (viii), asymptotic variance $v_{\sigma^2}^2$ of $\hat{\sigma}^2$ is equal to the lower right entry of $\Sigma_\theta$, and $v_{\sigma^2}^2 > 0$; under another additional condition (ix), the asymptotic covariance matrix $\Sigma_c$ of $\hat{c}$ is positive definite.

4. Under additional condition (vii), the matrix $\Sigma_\theta$ is positive definite.

*Proof of Statement 1.* The symmetric matrix $M_w := \overline{ww^T} - S$ converges in probability to the nonsingular matrix $\mu_2$ from condition (iv), therefore, $M_w$ is nonsingular with probability tending to 1 as $n \to \infty$. Hence the estimator $\hat{c}$ satisfies (8) with probability tending to 1 as $n \to \infty$. From this explicit formula we get directly that $\hat{c}$ is consistent, i.e., it converges in probability to $c$. Next, the estimator (10) converges in probability to $\sigma^2 > 0$, and the right-hand side of (10) is positive with probability tending to 1 as $n \to \infty$. Hence the estimator (14) coincides with (10) with probability tending to 1 as $n \to \infty$, and therefore, $\hat{\sigma}^2$ is consistent. Thus, the estimating equation (17) has a solution with probability tending to 1 as $n \to \infty$, with components of the estimating function given in (11) and (12).

The asymptotic normality of $\hat{\theta}$ is proven in a standard way, based on the equality that holds with probability tending to 1:

$$S(\hat{\theta}) = 0, \tag{A8}$$

where $S(\theta) := \sum_{i=1}^{n} s_\theta(y_i, w_i; \theta)$. By Taylor expansion of $S(\hat{\theta})$ in a neighborhood of the true value $\theta$, we obtain (remember that $\hat{\theta}$ is consistent):

$$S(\theta) + \frac{\partial S(\theta)}{\partial \theta^T}(\hat{\theta} - \theta) + r = 0, \tag{A8}$$

with

$$||r|| \leq O_P(1) \cdot ||\hat{\theta} - \theta||^2 = o_P(1) \cdot ||\hat{\theta} - \theta||. \tag{A9}$$

Hereafter $O_P(1)$ and $o_P(1)$ denote random sequences that are bounded in probability or converge to zero in probability, respectively. Next, by Central Limit Theorem (CLT) in Lyapunov form (see a convenient statement of CLT in Theorem 8 from [15]),

$$\frac{1}{\sqrt{n}} S(\theta) \to^d N(0, B), \tag{A10}$$

and by Law of Large Numbers

$$\frac{1}{n} \frac{\partial S(\theta)}{\partial \theta^T} \to^P A, \tag{A11}$$

where $\to^P$ denotes convergence in probability and matrices $A, B$ are given in (A6), moreover $A$ is nonsingular due to condition (iv). Now, relations (A8-A11) imply that $\sqrt{n} \cdot ||\hat{\theta} - \theta|| = O_P(1)$, hence $\sqrt{n} \cdot ||r|| = o_P(1)$. The convergence (A4-A6) follows from relations (A8), (A10), (A11) and Slutsky's Lemma (see Theorem 4.1 in [17]).

Notice that conditions (i-vi)) are used in particular to prove the convergence (A10), namely to show that the second limit in (A6) exists and to check the corresponding Lyapunov's condition.

*Proof of Statement 2.* The convergence in probability of both $\hat{A}$ and $\hat{B}$ follows from conditions (i-vii) and Law of Large Numbers. This implies the convergence of $\hat{\Sigma}_\theta$, because the matrix $A$ is nonsingular and $\hat{A}^+ \to^P A^{-1}$.

*Proof of Statement 3.* The matrix $A$ is block-diagonal, and $v_{\sigma^2}^2 = (A_{\sigma^2})^{-2} B_{\sigma^2} > 0$ under additional condition (viii). Next, under additional condition (ix), $\Sigma_c = A_c^{-1} B_c A_c^{-1}$ and it is positive definite.

*Proof of Statement 4.* Under additional condition (vii), the matrix (A8) is positive definite, because $B$ is positive definite.

This accomplishes the proof of Theorem.

## APPENDIX 2: P-VALUES FOR LARGE SAMPLE SIZE

Consider the model (2-3), but now for simplicity vector $x$ will be random (and not necessarily centered). Variables $x, e, \delta$ are presumed independent, with finite 4th moments. We suppose that conditions (i) and (ii) from section 2.1 are satisfied, except the following: matrix $S = \boldsymbol{Cov}(\delta)$ is now unknown. Denote $S_w = \boldsymbol{Cov}(w), S_x = \boldsymbol{Cov}(x)$; we presume that $S_x$ is nonsingular. Note that $S_w = S_x + S$, and $S_w$ is nonsingular as well.

Based on i.i.d. observed pairs given in (6), we want to compute p-value (36). Since $S = Cov(\delta)$ is unknown, we use certain presumed matrix $V_n$ instead. Because $S_w - S$ is positive definite, the $V_n$ will be selected as $V_n = \kappa \hat{S}_w$, $\kappa \in [0, 1)$, where $\hat{S}_w$ is sample covariance matrix of $w$,

$$\hat{S}_w = \frac{1}{n-1} \sum_1^n (w_i - \bar{w})(w_i - \bar{w})^T.$$

Here $\kappa$ regulates the level of the presumed errors in covariates. In what follows we suppose that $\kappa$ is nonrandom and does not depend on the sample size and has the same values for all x variables. Hence $V_n$ converges a.s. to $\kappa S_w$.

Now, we compute the estimator (7) as

$$\hat{c} = (\overline{ww^T} - V_n)^+ \overline{wy},$$

which converges a.s. to $\Lambda c$ with

$$\Lambda = (Eww^T - \kappa S_w)^{-1}(Exx^T). \tag{A12}$$

For the corresponding residual sum of squares $RSS$ (see Section 2.2) it holds

$$\frac{RSS}{n} \to^{P1} E(y - (\Lambda c)^T w)^2 = E(c^T(I_m - \Lambda^T)x + e - (\Lambda c)^T \delta)^2 =$$
$$= c^T(I_m - \Lambda^T) Exx^T (I_m - \Lambda)c + \sigma^2 + c^T \Lambda^T S \Lambda c.$$

Hereafter $\to^{P1}$ denotes the almost sure convergence as $n \to \infty$.

The statistic $T$ from (36) is a fraction; denote its numerator and denominator as $A_n$ and $B_n^{1/2}/\sqrt{n}$, respectively. We have

$$A_n = \frac{RSS}{n} - \sigma_0^2 - \hat{c}^T V_n \hat{c} \to^{P1} A_\infty,$$
$$A_\infty := c^T(I_m - \Lambda^T) Exx^T (I_m - \Lambda)c + c^T \Lambda^T (S - \kappa S_w) \Lambda c + (\sigma^2 - \sigma_0^2), \tag{A13}$$
$$B_n \to^{P1} B_\infty := E(y - (\Lambda c)^T w)^4 - (E(y - (\Lambda c)^T w)^2)^2 > 0.$$

Consequently, we obtain the following asymptotics for p-value (36):

$$\text{if } A_\infty > 0 \text{ then } T \to^{P1} +\infty \text{ and } p \to^{P1} 0, \tag{A14}$$
$$\text{if } A_\infty < 0 \text{ then } T \to^{P1} -\infty \text{ and } p \to^{P1} 1, \tag{A15}$$
$$\text{and if } A_\infty = 0 \text{ then } T \to^{P1} 0 \text{ and } p \to^{P1} \frac{1}{2}. \tag{A16}$$

In cases (A15-A16), the null hypothesis is *not rejected* for $n$ large enough (here $\sigma_0^2$ is quite large), and in case (A14), the null hypothesis (28) is *rejected* for $n$ large enough (here $\sigma_0^2$ is comparatively small). For finite sample, at each interval for $\kappa$ with $A_\infty \leq 0$, we have quite large p-values and the null hypothesis is *not rejected*.

Consider a particular case $c \neq 0$ and $\kappa = 0$ (i.e., it is presumed that there is no measurement errors), then

$$\Lambda = \Lambda_0 := (Eww^T)^{-1}(Exx^T), \quad I_m - \Lambda_0 = (Eww^T)^{-1} S,$$
$$A_\infty = c^T(I_m - \Lambda_0)^T Exx^T (I_m - \Lambda_0)c + c^T \Lambda_0^T S \Lambda_0 c + (\sigma^2 - \sigma_0^2).$$

If $\sigma_0^2 < \sigma^2$, then $A_\infty > 0$, $H_0$ is false and rejected for large $n$. If

$$\sigma_0^2 > \sigma^2 + c^T(I_m - \Lambda_0)^T Exx^T (I_m - \Lambda_0)c + c^T \Lambda_0^T S \Lambda_0 c,$$

then $A_\infty < 0$, $H_0$ is true and not rejected for large $n$.